\documentclass[prb,onecolumn,showpacs]{revtex4}

\usepackage{amssymb,amsmath,bm,epsfig,graphics,subfigure}
\newcommand{\tef}{transverse electric field}
\renewcommand{\bf}[1]{\mathbf{#1}}
\newcommand{\pd}[1]{\partial_{#1}}
\renewcommand{\d}{\text{d}}

\begin{document}

\title{Formation of subgap states in carbon nanotubes due to a local transverse electric field}
\author{Jesse M. Kinder}
	\email{kinder@physics.upenn.edu}
	\affiliation{Department of Physics, University of Pennsylvania\\Philadelphia, PA 19104}
\author{E.J. Mele}
	\affiliation{Department of Physics, University of Pennsylvania\\Philadelphia, PA 19104}
\date{\today}

\begin{abstract}

We introduce two simple models to study the effect of a spatially localized transverse electric field on the low-energy electronic structure of semiconducting carbon nanotubes. Starting from the Dirac Hamiltonian for the low energy states of a carbon nanotube, we use scattering theory to show that an arbitrarily weak field leads to the formation of localized electronic states inside the free nanotube band gap. We study the binding energy of these subgap states as a function of the range and strength of the electrostatic potential. When the range of the potential is held constant and the strength is varied, the binding energy shows crossover behavior: the states lie close to the free nanotube band edge until the potential exceeds a threshold value, after which the binding energy increases rapidly. When the potential strength is held constant and the range is varied, we find resonant behavior: the binding energy passes through a maximum as the range of the potential is increased. Large electric fields confined to a small region of the nanotube are required to create localized states far from the band edge.

\end{abstract}


\maketitle

Single-walled carbon nanotubes are cylindrical fullerenes with a radius on the order of a nanometer and lengths that can exceed a micrometer. They occur in both semiconducting and metallic species. The conductivity has a purely geometric origin that allows all nanotubes to be divided into three types: moderate-gap semiconductors, narrow gap semiconductors, and metals. The most common is the first type, which have a band gap between 0.1 eV and 1 eV that scales inversely with the tube radius $R$. This band gap is required in order for the electron wave function to be single-valued. The narrow-gap semiconductors have a band gap on the order of 0.01 eV that scales inversely with $R^2$ and arises from the distortion of the honeycomb lattice when it is wrapped into a cylinder. In the metallic armchair nanotubes, the formation of a band gap is prohibited by symmetry.\cite{kane97} We only consider moderate-gap semiconducting nanotubes in this paper.

In applications involving carbon nanotubes as electronic and optical components, a method for controllably and continuously varying the size of the intrinsic band gap would be useful. In principle, elastic deformations and magnetic flux through the a nanotube can be used to control the band gap.\cite{kane97,ando02} Another possibility is to tune the band gap with a static electric field perpendicular to the nanotube axis. A \tef{} leads to a Stark shift of the electronic energy levels. The scale of this shift is set by the electric potential difference across the diameter of the nanotube: $\Delta V \approx 2 E_0 R$, where $E_0$ is the magnitude of the electric field. In analogy with the Stark effect in atomic physics, one might expect a continuous reduction in the nanotube band gap as the strength of the electric field is increased. This is the case for boron-nitride nanotubes, which are wide-gap semiconductors with a band gap on the order of 5 eV. A Stark shift was predicted and has been observed.\cite{chen04, khoo04, ishi05}

Several authors have studied the effect of a uniform static transverse electric field on the band structure 
of moderate-gap single-walled carbon nanotubes.\cite{cho02, nov02, rot03, chen04, pach05, li06, gun06, nov06}  Instead of a linear or quadratic Stark shift in the band gap, numerical simulations indicate crossover behavior. The change in band gap is small until the electric field exceeds a threshold, after which the band gap decreases linearly with increasing field strength.\cite{cho02, chen04, pach05, li06}

Novikov and Levitov have used the low-energy Dirac theory for carbon nanotubes to provided a natural explanation of these numerical results.\cite{nov02,nov06} In the Dirac theory, the band gap is fixed at its zero-field value until the electric field exceeds a critical strength. Below the critical field, the band gap is protected by a chiral gauge symmetry; above the critical field, the effective mass of the charge carriers changes sign, and the band gap closes around the symmetry-protected point.

Numerical simulations do not exhibit critical behavior --- there are small increases or decreases in the band gap, even in very weak fields. However, changes in the band gap are small below a threshold field strength. As shown in App.~\ref{app:table}, the critical field predicted by Novikov and Levitov is smaller than the threshold fields reported in tight binding calculations. The work of Novikov and Levitov shows that the linear Dirac theory provides a qualitative description of the response of semicondcuting nanotubes to a uniform \tef: crossover behavior. This simple model captures the essential physical properties of a nanotube necessary to understand the general features its electronic response.

In this paper, we study a nanotube in a \emph{nonuniform} field and explore the possibility that spatially varying fields could couple to the minimum gap. The chiral symmetry of the uniform field model is broken by an inhomogeneous electric field, and the electronic response to a nonuniform field is qualitatively different.

Our work suggests the formation of exponentially localized subgap states is a general feature of the electronic response of semiconducting carbon nanotubes to \tef{}s localized to a small region along the tube axis. Since the states are localized, they do not reduce the band gap for free particle transport along the nanotube axis. This is different than the response to a uniform field, in which the gap for excitation from the valence band to the conduction band is modified. Our results indicate that this gap remains fixed in a field localized to a small region along the nanotube axis, even when the field strength is large.

\begin{figure}[hbt]
	\begin{center}
		\epsfig{file=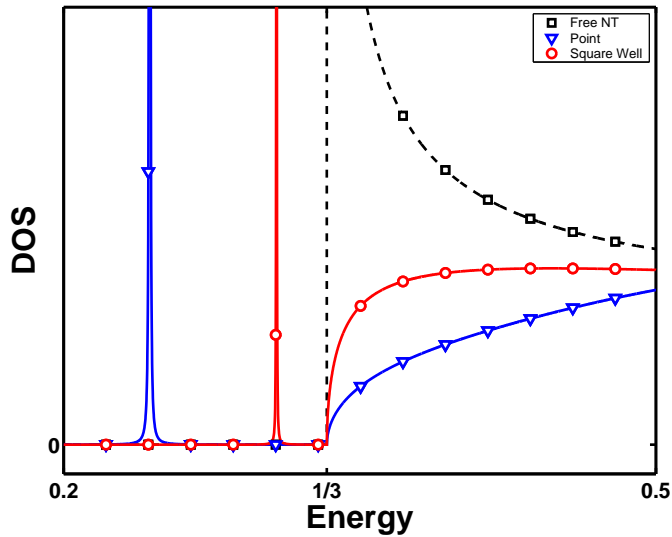, width=0.5\linewidth}
		\caption{(Color Online) The local density of states near the center of a carbon nanotube for energies close to the free nanotube band edge at $\varepsilon = 1/3$. Energies are given in units of $\hbar v_F/R$. The plot shows bound states inside the band gap and a modification of the density of states near the band edge.}
		\label{fig:dosPlot}
	\end{center}
\end{figure}

We study two simple models for a localized electric field. In both models, the potential modifies the density of states in two ways: 1) electron scattering leads to periodic variations in the density of states along the nanotube axis, and 2) the potential creates exponentially localized states inside the free nanotube band gap.  Fig.~\ref{fig:dosPlot} is a plot of the density of states near the band edge for a free nanotube and the two models we studied. The potentials define the center of an infinite nanotube. The plot shows the density of states at a distance of one tube radius from this origin.

Localized states form inside the free nanotube band gap in arbitrarily weak fields; however, the binding energy as a function of potential strength shows crossover behavior similar to that observed in numerical simulations of nanotubes in uniform \tef{}s. When the range of the potential is held constant, the subgap states lie close to the band edge until the electric fields exceeds a threshold strength. 

The remainder of the paper is divided into four sections. In Sec.~\ref{sec:method}, we present the Hamiltonian for a semiconducting carbon nanotube in a static \tef{} and outline the scattering theory we used to study the spectra of these systems. Next, we analyze two models models in which the potential due to the applied field is localized along the nanotube axis. In Sec.~\ref{sec:delta}, we consider an electric field that can be modeled as a delta function in the long-wavelength Hamiltonian. In Sec.~\ref{sec:square}, we consider an effective square well potential. In Sec.~\ref{sec:final}, we compare the solutions of these models. Our main results are presented in Figs.~\ref{fig:delta}, \ref{fig:sqStrength}, and \ref{fig:sqSize}.

\section{Hamiltonian and Scattering Theory}\label{sec:method}

The electronic structure of a carbon nanotube is inherited from the band structure of two-dimensional graphene, a planar honeycomb lattice of carbon atoms. Graphene is a semimetal with a vanishing band gap at the six corners of the first Brillouin zone, the $\bf{K}$ and $\bf{K}'$ points.\cite{wal47} Since the honeycomb lattice has a two atom basis, the Bloch wave functions are a sum of amplitudes for each sublattice, which we denote $A$ and $B$:
\begin{equation}
	\Psi_\bf{k} (\bf{r}) = \psi_{A,\bf{k}} (\bf{r}) + \psi_{B,\bf{k}} (\bf{r}).
\end{equation}

Near the $\bf{K}$ and $\bf{K}'$ points, the electron wave function can be approximated by introducing envelope functions to describe long-wavelength variations of the exact wave functions at $\bf{K}$ or $\bf{K}'$. Thus, if the total wave vector is $\bf{k} = \bf{K} + \bf{q}$, then
\begin{equation}
	\Psi_\bf{k} (\bf{r}) = \psi_{A,\bf{K}} (\bf{r}) \, v_\bf{K} (\bf{r}) + \psi_{B,\bf{K}} (\bf{r}) \, w_\bf{K} (\bf{r}). \label{eq:envelope}
\end{equation}
The Bloch functions $\psi_{\alpha}$ are the exact $\bf{K}$ wave functions and oscillate on the scale of the lattice spacing. The envelope functions $v$ and $w$ oscillate on much longer length scales.

When expanded to linear order in $\bf{q}$, the resulting effective Hamiltonian for the envelope functions near the $\bf{K}$ point\footnote{The corresponding effective Hamiltonian for envelope functions near the $\bf{K}'$ points is the time reverse of the Hamiltonian at $\bf{K}$.} is a massless Dirac equation:\cite{sem84, mele84,kane97}
\begin{equation}
	-i \hbar v_F \, \left( \sigma_x \pd{x} + \sigma_y \pd{y} \right) \begin{pmatrix} v_\bf{K} (\bf{r}) \\ w_\bf{K} (\bf{r}) \end{pmatrix} = \mathcal{E}(\mathbf{q}) \, \begin{pmatrix} v_\bf{K} (\bf{r}) \\ w_\bf{K} (\bf{r}) \end{pmatrix}. \label{eq:graphene}
\end{equation}
The factor $\hbar v_F$ comes from the tight-binding model of graphene. Its value is $ 3 a t / 2 \approx 0.53 \, \text{eV} \cdot \text{nm}$ where $t$ is the nearest neighbor hopping energy and $a$ is the distance between neighboring atoms. The effective Hamiltonian describes spinless electrons in the graphene lattice. The two components of the pseudospinor give the relative amplitude for an electron or hole to be localized on the $A$ and $B$ sublattices.

The low-energy theory of a carbon nanotube follows from this Hamiltonian. One defines a circumference vector $\bf{C}$ in the graphene lattice, which determines both the radius $R$ and chiral angle $\theta$ of the nanotube. The full electron wave function must be single-valued on the surface of the nanotube: $\Psi_\bf{k} ( \bf{r} + \bf{C} ) = \Psi_\bf{k} (\bf{r})$. This requirement collapses the energy surface of graphene into a set of discrete energy bands. If none of the energy bands passes through the $\bf{K}$ and $\bf{K}'$ points, then the nanotube will have a band gap inversely proportional to its radius, and the envelope functions $v$ and $w$ satisfy quasiperiodic boundary conditions.\cite{kane97} These are the moderate gap semiconducting carbon nanotubes we consider in this paper.

The natural description of a nanotube is in cylindrical coordinates. The displacement along the nanotube axis, $z$, and angular displacement around the circumference, $\phi$, are related to the $(x,y)$ coordinates of the graphene sheet by a rotation through the chiral angle $\theta$. (We define $\theta$ to be the angle between the $\bf{K}$ point and the nanotube axis, so $\theta = 0$ for armchair nanotubes and $\theta = \pi / 6$ for zig-zag nanotubes.) The transformations are
\begin{equation}
	\begin{pmatrix} z \\ R \phi \end{pmatrix} = \begin{pmatrix} x \cos \theta + y \sin \theta \\ y \cos \theta - x \sin \theta \end{pmatrix}.
\end{equation}

In the coordinates of the nanotube, Eq.~(\ref{eq:graphene}) can be expressed
\begin{equation}
	\mathcal{H} = -i ( \hbar v_F / R ) \, e^{-i \sigma_z \theta / 2 } \, \left( R \sigma_x \pd{z} + \sigma_y \pd{\phi} \right) \, e^{ i \sigma_z \theta / 2 }. \label{eq:freeNT}
\end{equation}

The energy spectrum of this free nanotube Hamiltonian is a series of hyperbolic bands given by
\begin{equation}
	\mathcal{E}_m(q) = \pm ( \hbar v_F / R ) \sqrt{ (qR)^2 + \Delta_m^2 }, \label{eq:spectrum}
\end{equation}
where $q$ is momentum along the nanotube axis and $\Delta_m = m + \nu/3$. The integer $m$ is the subband index, and $\nu = \pm 1$ is required by the quasi-periodic boundary conditions on the envelope functions. Without loss of generality, we choose $\nu = 1$ in the remainder of the paper.

Placing the nanotube in a static electric field perpendicular to its axis introduces the scalar potential
\begin{equation}
	V(z,\phi) = e \, ( E_0 / \kappa ) \, R f(z) \cos \phi. \label{eq:potential}
\end{equation}
$E_0$ sets the scale of the applied field strength; variations in field strength along the nanotube axis are described by the dimensionless function $f(z)$. The electric field that enters the Hamiltonian is the screened electric field, not the applied electric field. Calculations indicate the effects of depolarization are significant and reduce the strength of the external field by a factor of roughly 5.\cite{ben95,nov02} We include depolarization effects through the constant $\kappa$.

The natural energy scale of the nanotube system is $\hbar v_F / R$. We will use this quantity as the unit of energy and work only with the reduced Hamiltonian. The reduced potential strength is $u = e E_0 R^2 / \kappa \hbar v_F$. This is a convenient parameter to use in calculations, but it obscures the magnitude of the fields involved. For reference, $u=1$ in a nanotube with a radius of 1 nm corresponds to an applied electric field strength of about 10 V/nm.

The reduced Hamiltonian --- with dimensionless eigenvalues $\varepsilon$ --- is the starting point for both of our models:
\begin{equation}
	\mathcal{H} = -i e^{-i \sigma_z \theta / 2 } \, \left( R \sigma_x \pd{z} + \sigma_y \pd{\phi} \right) \, e^{ i \sigma_z \theta / 2 } + u \, f(z) \, \cos \phi. \label{eq:hRed0}
\end{equation}

Because the potential depends on the azimuthal angle $\phi$, angular momentum is no longer a good quantum number. However, the potential is single-valued around the circumference of the nanotube. The Hamiltonian has a discrete symmetry under $\phi \rightarrow \phi + 2\pi$, so its eigenstates can be expanded in the basis of angular momentum eigenstates of the unperturbed Hamiltonian. The potential is proportional to $\cos \phi$, so it will only connect states whose angular momenta differ by $\pm 1$: i.e., electronic states in neighboring energy bands. Projecting out the azimuthal degree of freedom in the Hamiltonian of Eq.~(\ref{eq:hRed0}) gives
\begin{equation}
	\mathcal{H}_{m,n} = e^{-i \sigma_z \theta / 2 } \, \left( \sigma_y \Delta_m - i R \sigma_x \pd{z}  \right) \, e^{ i \sigma_z \theta / 2 } \, \delta_{m,n} + (u / 2 ) \, f(z) \, \delta_{m,n\pm1}, \label{eq:hRed}
\end{equation}
where $m$ and $n$ are the band indices of the unperturbed nanotube.

We study this Hamiltonian for two different models of a nonuniform \tef. In the first, we treat the potential as a point scatterer with $f(z) \propto \delta(z)$; in the second, we model the potential as a square well so that $f(z)$ is constant within a region of length $L$ and zero outside. Neither of these models can represent a real electric field because neither potential satisfies Maxwell's equations. However, common features in both these models suggests a universal response to a localized static transverse electric fields: the formation of localized states with energies inside the unperturbed band gap.

Eq.~(\ref{eq:hRed}) is an effective Hamiltonian for the long-wavelength modes of the carbon nanotube. Accordingly, $f(z)$ is an effective profile of the electric field. The physical interpretation of $f(z)$ can be made precise by considering the three important length scales in the problem: the length over which the potential changes significantly, $\Delta x$; the wavelength (or decay length) of the envelope functions, $\lambda = 2 \pi / |q|$; and the distance between neighboring atoms of the honeycomb lattice, $a$. The natural cutoff for the linear Dirac theory of electrons in carbon nanotubes is $\lambda \sim R$.\footnote{The linear Dirac theory is the first term in an expansion of the tight binding Hamiltonian in powers $qa$ around one of its minima. Imposing the cutoff $qR \alt 1$ ensures that $qa \ll 1$, so that higher order terms in the expansion such as the trigonal warping term may be ignored.} Particles with energies near the band edge may have wavelengths much larger than the radius, on the order of the tube length. As a result, we may consider the wavelength of the envelope functions to be bounded from below: $\lambda \agt R$.

\begin{figure}[htb]
\centering
	\subfigure[Point Potential]{\epsfig{figure=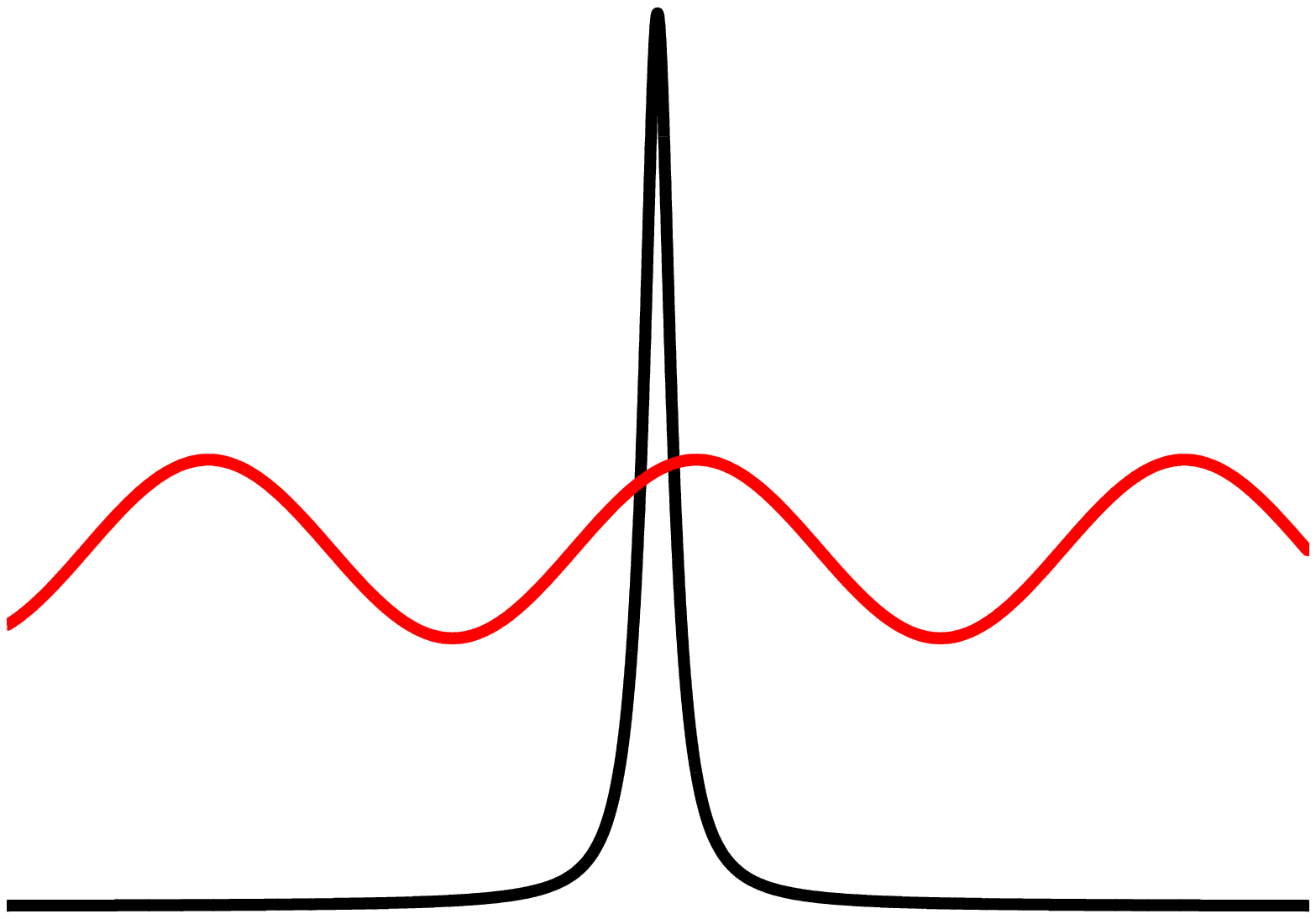,width=0.4\textwidth}}\qquad \qquad
	\subfigure[Square Well]{\epsfig{figure=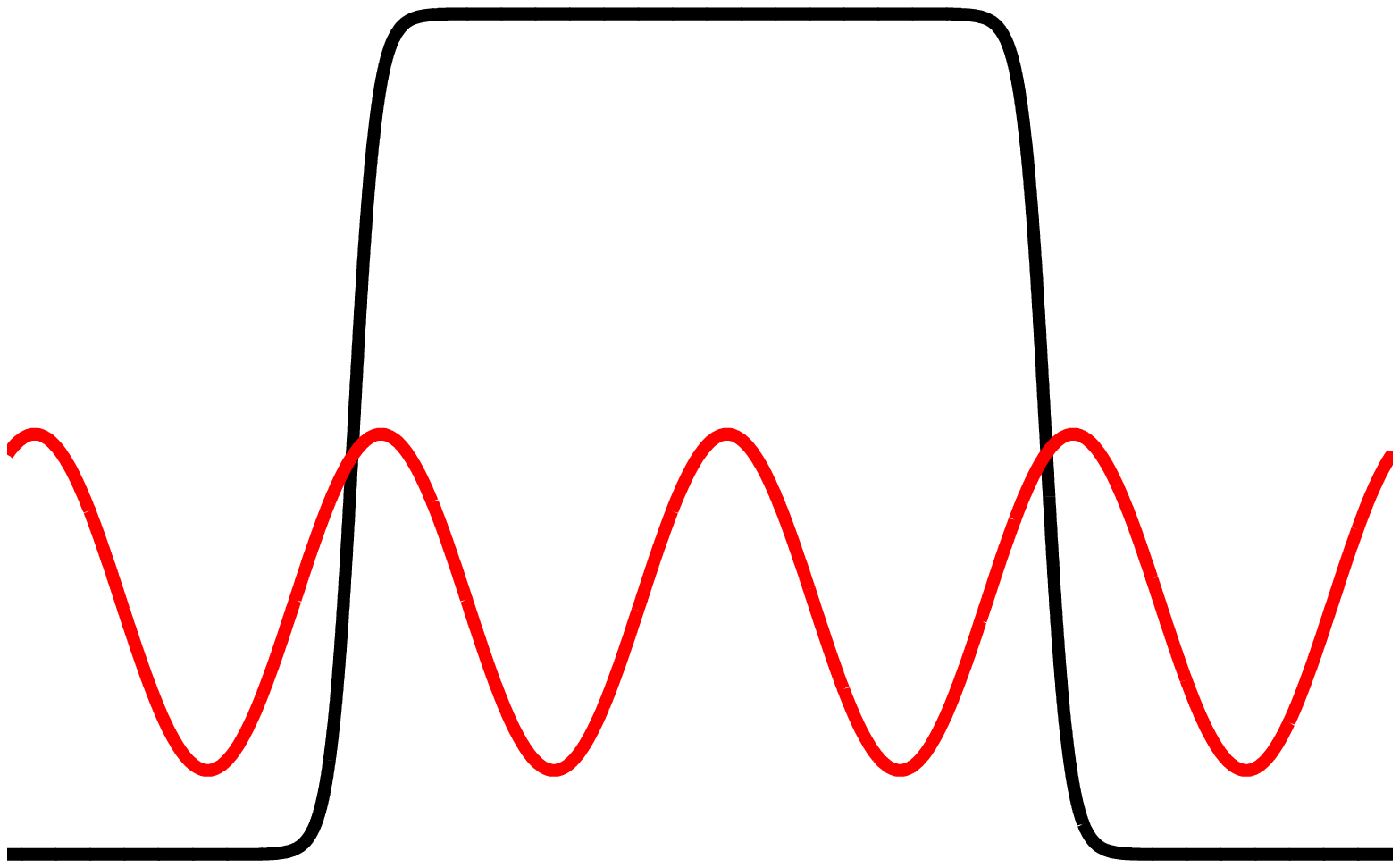,width=0.4\textwidth}}
\caption{(Color Online) Field profiles that can be analyzed with the models presented in this paper. The black lines show the electric field strength along the nanotube axis. The red lines indicate spatial variations in the envelope functions along the axis. The potentials are roughly constant on length scales comparable to the interatomic spacing, but change significantly on the scale of the envelope function wavelength.\\In (a), the potential rises from zero to its maximum value and falls back to zero within a single wavelength of the envelope function. In the effective Hamiltonian, this potential can be modeled as a point scatterer. In (b), the potential rises from zero to its maximum value over a distance smaller than the envelope function wavelength. It is constant for several wavelengths, then falls back to zero over a distance smaller than the wavelength. In the effective Hamiltonian, it can be represented by a square well.}
\label{fig:potentials}
\end{figure}

We consider potentials that satisfy $a \ll \Delta x \ll \lambda$. The physical field is continuous. The change in potential over distances comparable to the lattice spacing is negligible, but on length scales comparable to the wavelength of the envelope functions, the field strength changes abruptly. Since the physical potential is roughly constant over distances comparable to the lattice spacing, there is no intervalley scattering between the $\bf{K}$ and $\bf{K}'$ points even though the potential in the effective Hamiltonian is discontinuous. For such a potential, the energy spectrum can be determined by considering a single valley in isolation. Fig.~\ref{fig:potentials} illustrates the types of potentials to which the simple models discussed in this paper might apply.

In the remainder of this paper, we will ignore the chiral angle $\theta$. The dependence of $\mathcal{H}_{m,n}$ on $\theta$ can be removed with a unitary transformation. This is discussed in App.~\ref{app:noChiral}, in which we show that the density of states and bound state energies of Eq.~(\ref{eq:hRed}) do not depend on the chiral angle. The results derived for $\theta = 0$ apply to chiral tubes as well.\footnote{Including the trigonal warping term in Eq.~(\ref{eq:hRed}) introduces a dependence on the chiral angle that cannot be removed with a unitary transformation, but we ignore this effect by restricting ourselves to small wave vectors along the nanotube axis, as discussed in the preceding footnote.}

To study the effects of these potentials on the electronic structure of a carbon nanotube, we use scattering theory to determine the Green function of Eq.~(\ref{eq:hRed}) from the Green functions of a free nanotube. The singularities of the Green function give the energy spectrum, and its trace gives the local density of states. Free nanotube Green functions are shown in an open font: $\mathbb{G}$. The full Green functions for a nanotube in the applied field are displayed in a script font: $\mathfrak{S}$.

The Green function for a charge carrier in band $m$ of a free nanotube that satisfies outgoing boundary conditions as $|z - z'|$ approaches $\pm \infty$ is
\begin{equation}
	\mathbb{G}_{m} (\varepsilon,z,z') = \dfrac{ \varepsilon + \sigma_y \Delta_m + \mbox{sgn}(z-z') \, \sigma_x \xi_m(\varepsilon) }{ 2 i R \xi_m(\varepsilon)  } \, e^{ i \xi_m(\varepsilon) \, |z-z'| / R  } \label{eq:green}
\end{equation}
where $\xi_m(\varepsilon) = \sqrt{ \varepsilon^2 - \Delta_{m}^{2} }$. This expression is derived in App.~\ref{app:gFunc}. The Green functions are $2 \times 2$ matrices to account for the psuedospin of the envelope functions.\footnote{Multiplication by the $2 \times 2$ identity matrix is implied for scalars appearing in expressions that also involve the Pauli matrices.} When $|\varepsilon| < \Delta_m$, $\xi_m(\varepsilon)$ is imaginary and the Green functions decay exponentially as $|z-z'|$ increases.

The Green functions of the perturbed system are solutions of Dyson's equation: $ \mathfrak{S} = \mathbb{G} + \mathbb{G} \, V \, \mathfrak{S}$. Formally, the solution is
\begin{equation}
	\mathfrak{S} = \dfrac{ \openone }{ \openone - \mathbb{G} V } \, \mathbb{G}, \label{eq:formal}
\end{equation}
where $\openone$ is the $2 \times 2$ identity matrix. Expanding this expression gives an infinite series:
\begin{equation}
	\mathfrak{S} = \mathbb{G} + \mathbb{G} V \left\{ \sum_{n=0}^{\infty} (\mathbb{G} V)^n \right\} \mathbb{G}. \label{eq:series}
\end{equation}
This is an operator equation, and products are shorthand for integration over position along the nanotube axis and summation over all bands. In general, this equation cannot be solved exactly. We make a series of approximations that allows us to carry out a partial summation and approximate the exact Green function.

As shown in App.~\ref{app:neighbors}, when the energy of a charge carrier is close to the band edge, the largest terms in the Dyson series are those that only contain transitions to the neighboring bands. Retaining only Green functions from the first three bands ($m = 0, \pm1$), Eq.~(\ref{eq:series}) can be summed to give the full Green function for the lowest band:
\begin{equation}
	\mathfrak{S}_0 = \mathbb{G}_0 + \mathbb{G}_0 \Gamma_0 \dfrac{ \openone }{ \openone -  \mathbb{G}_0 \Gamma_0} \mathbb{G}_0 . \label{eq:genSol0}
\end{equation}
$\Gamma_0 = V (\mathbb{G}_{+1} + \mathbb{G}_{-1} ) V$ is an effective potential for the lowest band. It will be nonlocal in general; however, for the models studied in Sec.~\ref{sec:delta} and \ref{sec:square}, the effective potential only depends on energy. The general form of the full Green function is then
\begin{equation}
	\mathfrak{S}_0 (\varepsilon,z,z') = \mathbb{G}_0(\varepsilon,z,z') + \mathbb{G}_0(\varepsilon,z,0) \, \Gamma_0(\varepsilon) \, \dfrac{ \openone }{ \openone - \mathbb{G}_0(\varepsilon,0,0) \, \Gamma_0(\varepsilon) + i \eta } \, \mathbb{G}_0(\varepsilon,0,z'). \label{eq:genSol}
\end{equation}
A small imaginary number, $i \eta$, has been included in the denominator of Eq.~(\ref{eq:genSol}) so the poles in the complex energy plane correspond to physical particles. This is the retarded Green function.

The density of states in the nanotube is proportional to the trace of the Green function. The poles of $\mathfrak{S}_0$ are bound state energies, and branch cuts correspond to energy bands.\cite{matt92,don92} When the energy approaches a band edge of the unperturbed Hamiltonian, either $\mathbb{G}_0$ or $\Gamma_0$ diverges. In either case, $\mathfrak{S}_0$ remains finite, which means $\mathfrak{S}_0$ does not share any singularities of the free nanotube Green functions. (This is clear in Fig.~\ref{fig:dosPlot}. The free nanotube density of states has a van Hove singularity at the band edge, while the density of states vanishes at the band edge in the presence of a potential.)

If there are any poles in Eq.~(\ref{eq:genSol}), they occur when the determinant of the denominator vanishes: i.e.,
\begin{equation}
	\det [ \openone - \mathbb{G}_0(\varepsilon,0,0) \, \Gamma_0(\varepsilon) ] = 0 .\label{eq:secular}
\end{equation}
This equation yields a sixth-order polynomial, and we were unable to find a general solution for energy as a function of potential strength. To find the roots, we created a simple numerical routine to determine the energy  (within a user-specified tolerance) at which the determinant changes sign.

In both of the models, numerical results suggest that two localized states are created inside the band gap for any nonzero potential --- one for the valence band and one for the conduction band. In App.~\ref{app:noCrit}, we show analytically that the delta function model has states inside the free nanotube band gap for arbitrarily weak fields. The analog of this proof for the square well model is discussed in Sec.~\ref{sec:square}. 

The fact that bound states come in pairs reflects the symmetry of the particle and hole spectra for the potential in Eq.~(\ref{eq:potential}). When the charge of the particle is reversed, $V(z,\phi) \rightarrow -V(z,\phi)$, which is equivalent to physically rotating the nanotube by $\pi$ about its axis. Since the point at which $\phi = 0$ is arbitrary, the particle and hole spectra must be identical, and all energy eigenvalues of Eq.~(\ref{eq:hRed}) --- including bound state energies --- come in pairs $\pm \varepsilon$.

In the more familiar case of a charged impurity, an attractive potential for electrons is repulsive for holes, and the bound state spectrum is not symmetric. The difference between the spectra of a charged impurity and the potential considered in this paper arises from the angular dependence of the latter.

We define $\varepsilon_B$ to be the (non-negative) energy at which the full Green function has a pole. To study the dependence of the bound state energies on the range and strength of the potential, we plot the binding energy $\gamma = 1 - \varepsilon_B / \Delta_0$, which is the normalized difference in energy between the localized state and the free nanotube band edge. Our numerical results show $0 < \gamma \leq 1$. If $\gamma = 0$, there are no states inside the gap; if $\gamma = 1$, the subgap states lie on top of each other at the center of the free nanotube band gap.

In the following two sections, we apply the general formalism described here to two specific models. In the first, we treat the effective field profile as a delta function. In the second, we model the effective field profile as a square well. We study the binding energy of the subgap states as a function of the potential size $L$ and strength $u$. Figs.~\ref{fig:delta}, \ref{fig:sqStrength}, and \ref{fig:sqSize} illustrate our results.

\section{Delta Function Model}\label{sec:delta}

In this section, we study the electronic structure of a carbon nanotube in a \tef{} that can be treated as a delta function in the effective Hamiltonian. (See Fig.~\ref{fig:potentials}(a).) 

The physical potential can be written $V(z) = V_0 \, h(z)$, where $h(z)$ is some function that is localized to a small region of the nanotube axis --- for instance, a Lorentzian distribution peaked around $z = 0$. We define the range of the potential to be $L = \int \d z \, h(z)$, then replace $f(z)$ by $L \, \delta(z)$ in Eq.~(\ref{eq:hRed}). This ensures that the integrated strength of the potential is unchanged. A delta function potential in the effective Hamiltonian will be a reasonable approximation as long as $L \alt R$.

The effective Hamiltonian for the delta function model is
\begin{equation}
	\mathcal{H}_{m,n} = \left( \sigma_y \Delta_m - i R \sigma_x \pd{z}  \right) \, \delta_{m,n} + ( u L / 2 ) \, \delta(z) \, \delta_{m,n\pm1}. \label{eq:hDelta}
\end{equation}
Since the potential is proportional to $uL$, the range $L$ and strength $u$ of the potential cannot be varied independently. Any particular choice of $(u,L)$ is equivalent to an infinite class of choices $(u \beta , L / \beta)$.

When the potential is proportional to a delta function, the integrals of Eq.~(\ref{eq:series}) can be evaluated exactly. For instance, the term $\mathbb{G}_0 V \mathbb{G}_{\pm 1} V \mathbb{G}_0$ is
\begin{equation}
	(uL/2)^2 \, \mathbb{G}_0 (\varepsilon, z, 0 ) \mathbb{G}_{\pm 1}(\varepsilon,0,0) \mathbb{G}_0 (\varepsilon, 0, z' ).
\end{equation}
As a result, the full Green function is given by Eq.~(\ref{eq:genSol}) with
\begin{equation}
	\Gamma_0(\varepsilon) = (uL/2)^2 [ \mathbb{G}_{+1}(\varepsilon,0,0) + \mathbb{G}_{-1}(\varepsilon,0,0) ]. \label{eq:gamma}
\end{equation}

The local density of states is proportional to the imaginary part of the trace of Eq.~(\ref{eq:genSol}). When $\varepsilon$ lies inside the free nanotube band gap, the Green functions are exponentially decaying functions, as mentioned in Sec.~\ref{sec:method}. Since the bound states lie inside the free nanotube band gap, the local density of states decays exponentially with increasing distance from the potential:
\begin{equation}
	\rho(\varepsilon_B,z) \propto \exp [  ( 2|z| / R ) \sqrt{ \Delta_0^2 - \varepsilon_B^2 } ]. \label{eq:dos}
\end{equation}
The bound state wave functions are centered on the potential and decay exponentially, like the solutions of the 1D Schr\"{o}dinger equation with an attractive delta function potential.

\begin{figure}[hbt]	\begin{center}
		\epsfig{file=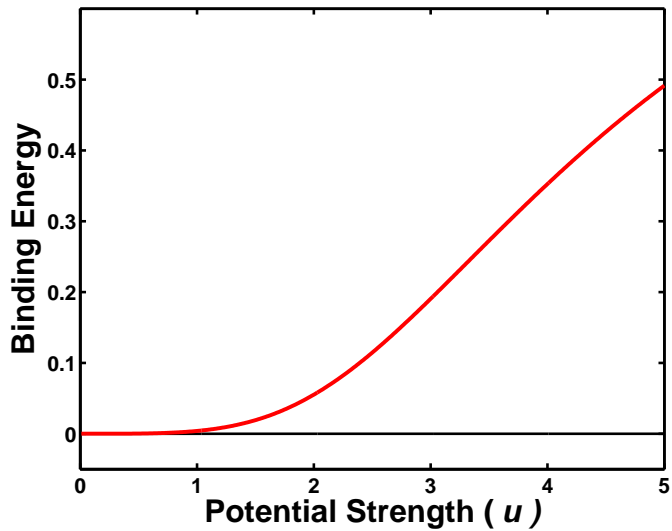, width=0.5\linewidth}
		\caption{(Color Online) Binding energy as a function of potential strength. The horizontal axis shows the value of the dimensionless parameter $u = e E_0 L / ( \hbar v_F / R )$ that characterizes the applied field strength. The line is $\gamma = 1 - \varepsilon_B / \Delta_0$ for $L = R$. The subgap states approach the middle of the gap as the potential strength is increased, but lie close to the band edge in weak fields.}
		\label{fig:delta}
	\end{center}
\end{figure}

Fig.~\ref{fig:delta} shows $\gamma = 1 - \varepsilon_B / \Delta_0$ as a function of potential strength $u$ when $L = R$. The binding energy shows crossover behavior similar to the band gap in a uniform \tef{}. Although the subgap states are very near the band edge when the field is weak, there is no critical behavior like that found by Novikov and Levitov. In App.~\ref{app:noCrit}, we show analytically that the Green function in Eq.~(\ref{eq:genSol}) has two poles for an arbitrarily weak potential and that the binding energy of the subgap states increases as $u^4$ for small $u$.

This quartic scaling is not unique to the Dirac model and occurs in nonrelativistic models as well. If the potential of Eq.~(\ref{eq:potential}) is introduced to a Schr\"{o}dinger equation describing electrons confined to the surface of a cylinder, the binding energies also grow with $u^4$. Because the potential only connects states in neighboring bands, the effective potential in a particular band is proportional to $u^2 \delta(z)$. The binding energy will scale with the square of the effective potential strength, which explains the $u^4$ dependence in Fig.~\ref{fig:delta}.

The figure shows that the potential must be strong ($ e \mathcal{E}_0 L \agt \hbar v_F / R$) to produce subgap states that lie more than a few percent below the band edge.

\section{Square Well Model}\label{sec:square}

In this section, we model the potential as a square well in the effective Hamiltonian. (See Fig.~\ref{fig:potentials}(b).) In this model, the strength and size of the potential can be varied independently, and the restrictions on the range of the potential are relaxed. We find the effect of changing the range of the potential is different than changing the field strength.

A square well potential in the effective Hamiltonian of Eq.~(\ref{eq:hRed}) corresponds to a physical field that increases from zero to its maximum strength in a distance that is small compared to the scale on which the envelope functions vary. The field strength is assumed to be constant for some length $L$, after which it decreases back to zero.

For the square well model to be a good approximation of the physical potential, $L$ must be smaller than the length of the nanotube. Otherwise, the electric field should be treated as uniform. There are no other physical restrictions on $L$. However, we are only able to solve Dyson's equation after imposing the additional restriction that $qL \ll 1$ where $q$ is the wave vector of the lowest band ($m=0$). Since we study states near the band edge where $q$ approaches zero, this is not a restrictive assumption. It makes the effective potential a point scatterer for states in the lowest band. The difference between this model and the delta function model is that the (exponentially decaying) states in neighboring bands ($m = \pm1$) interact with the potential in a region of finite size.

The effective Hamiltonian for the square well model is
\begin{equation}
	\mathcal{H}_{m,n} = ( \sigma_y \Delta_m - i R \sigma_x \pd{z}  ) \, \delta_{m,n} + ( u / 2 ) \, \Theta(z) \, \Theta(z-L) \delta_{m,n\pm1}, \label{eq:hSquare}
\end{equation}
where $\Theta(z)$ is the step function.

In the delta function model, $\delta(z)$ set all of the nonanalytic functions of the Green functions in Eq.~(\ref{eq:green}) equal to zero in the integrals of the Dyson series. For the square well model, the integral has to be evaluated piecewise to account for the terms in the Green function proportional to $\text{sgn}(z-z')$ and $|z-z'|$. For instance, the second-order term
\begin{equation}
	\mathbb{G}_0 V \mathbb{G}_{\pm1} V \mathbb{G}_0 = (u/2)^2 \int_{0}^{L} \d x \, \int_{0}^{L} \d y \, \mathbb{G}_0(\varepsilon,z,x) \mathbb{G}_{\pm1}(\varepsilon,x,y) \mathbb{G}_0(\varepsilon,y,z') \label{eq:integral}
\end{equation}
must be divided into eight regions of integration if both $z$ and $z'$ lie inside the square well. We are unable to derive a general formula for the $\mathcal{O}(u^{2n})$ terms of the Dyson series. As a result, we cannot sum the series, calculate the full Green function, find the density of states, or search for localized states below the band gap.

By requiring that $qL \ll 1$ where $q$ is the wave vector of the lowest band, we can approximate the integrals and sum the series of Eq.~(\ref{eq:series}). This approximation simplifies the expression for $\mathbb{G}_0(\varepsilon,z,z')$ in two ways. First, if $L \agt R$, then $qR$ is smaller than $qL$, which is negligible. Thus, we can ignore the term in the Green function proportional to $\text{sgn}(z-z')$. Second, when $\mathbb{G}_0$ appears inside an integral, its phase, $\exp ( i q |x - y| )$, can be replaced by a more convenient function. If $x$ and $y$ lie inside the region of integration and $z$ is arbitrary, the error introduced by the following substitutions is of order $qL$:
\begin{align}
	e^{i q |z-x|} & \longrightarrow e^{i q |z|} e^{i q x}, \notag \\
	e^{i q |x-y|} & \longrightarrow e^{i q (x + y)}. \label{eq:approx}
\end{align}
(Since $z$ may lie anywhere along the length of the nanotube, we do not assume $q|z|$ is small.) As an example, consider the following:
\begin{equation}
	\int_{0}^{L} \int_{0}^{L} \text{d}x \, \text{d}y \  e^{iq|x-y|} = \int_{0}^{L} \int_{0}^{L} \text{d}x \, \text{d}y \ e^{iq(x+y)} + \mathcal{O}(qL).
\end{equation}
Since we are interested in the limit $qL \ll 1$, the error is insignificant.

Assuming $qL \ll 1$, the integral in Eq.~(\ref{eq:integral}) gives
\begin{equation}
	\mathbb{G}_0 V \mathbb{G}_{\pm 1} V \mathbb{G}_0 = u^2 \Lambda_{\pm 1}(\varepsilon) \, \mathbb{G}_0 (\varepsilon, z, 0 ) \mathbb{G}_{\pm1}(\varepsilon,0,0) \mathbb{G}_0 (\varepsilon, 0, z' ) + \mathcal{O}(qL) ,
\end{equation}
The scalar function $\Lambda_{\pm 1}(\varepsilon)$ depends on $\varepsilon$ and $L$.
\begin{equation}
	\Lambda_{\alpha}(\varepsilon) = \frac{ R^2 }{ 2 ( \Delta_{\alpha}^{2} - \Delta_{0}^{2} ) } \, \left( x_\alpha(\varepsilon) + e^{-x_\alpha(\varepsilon)} - 1 \right), \label{eq:lambda}
\end{equation}
where $x_\alpha \equiv ( L / R ) \sqrt{ \Delta_{\alpha}^{2} - \varepsilon^2  } = i \xi_\alpha(\varepsilon) \, (L/R)$.

Ignoring the $\mathcal{O}(qL)$ terms, , all the integrations and summations can be carried out as in the delta function model and the full Green function is given by Eq.~(\ref{eq:genSol}) with
\begin{equation}
	\Gamma_0(\varepsilon) = u^2 \left[ \Lambda_{+1}(\varepsilon) \, \mathbb{G}_{+1} (\varepsilon,0,0) + \Lambda_{-1}(\varepsilon) \, \mathbb{G}_{-}(\varepsilon,0,0) \right].
\end{equation}

Comparing this expression with Eq.~(\ref{eq:gamma}) shows that when $qL \ll 1$, the square well model is equivalent to a delta function model with an energy-dependent range: $ \ell_\alpha (\varepsilon) = 2 \sqrt{\Lambda_\alpha (\varepsilon)}$. As a result, the arguments of App.~\ref{app:noCrit} also apply to the square well model.

When $L \ll R$ and $\varepsilon \approx \Delta_0$, $\Lambda(\varepsilon)$ approaches $L^2/4$, and the square well model reduces to the delta function model, for which there are bound states at any field strength. When $L \gg R$, the exponential term in Eq.~(\ref{eq:lambda}) can be ignored. An analysis similar to that of App.~\ref{app:noCrit} shows that there are bound states for arbitrarily weak fields in this limit as well. Since $\Lambda_\alpha(\varepsilon)$ is a smoothly increasing function of $L$, we interpolate between these two limits and conclude that a square well potential of any strength and range leads to the formation of localized states inside the free nanotube band gap.

As in the delta function model, the bound state wave functions decay exponentially away from the potential , and the binding energy is proportional to $u^4$ for small $u$.

\begin{figure}[hbt]
	\begin{center}
		\epsfig{file=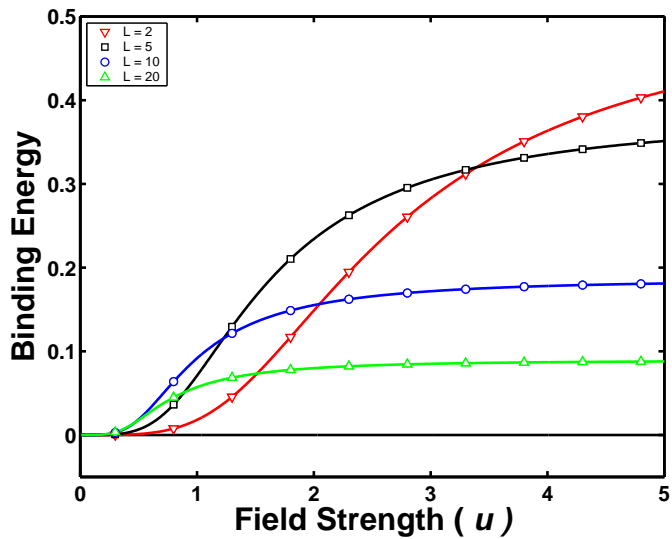, width=0.5\linewidth}
		\caption{(Color Online) Binding energy as a function of potential strength in the square well model when the width of the potential is fixed. The horizontal axis shows the value of the dimensionless parameter $u = e E_0 L / ( \hbar v_F / R )$ that characterizes the applied field strength. The lines are $\gamma = 1 - \varepsilon_B / \Delta_0$ for the potential widths $L/R$ shown in the legend. The shape of the individual curves is similar to the binding energy curve of the delta function model in Fig.~\ref{fig:delta}. Increasing the width of the well decreases maximum binding energy.}
		\label{fig:sqStrength}
	\end{center}
\end{figure}

Fig.~\ref{fig:sqStrength} shows the binding energy of the subgap states as a function of electric field strength when the range of the potential is fixed (fixed $L$, variable $u$). The shapes of the individual curves are similar to that of the delta function model in Fig.~\ref{fig:delta}. The range of the potential determines the asymptotic value of the binding energy as the field increases.

\begin{figure}[hbt]
	\begin{center}
		\epsfig{file=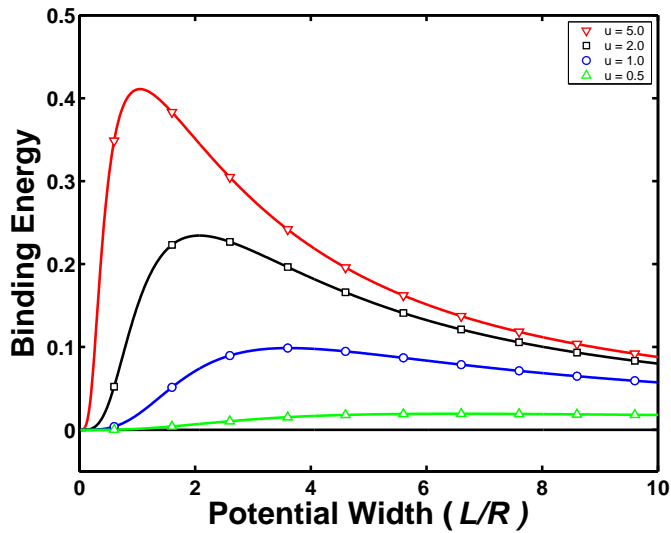, width=0.5\linewidth}
		\caption{(Color Online) Binding energy as a function of potential width in the square well model when the field strength is fixed. The horizontal axis gives the width of the square well, $L/R$. The lines are $\gamma = 1 - \varepsilon_B / \Delta_0$ for the potential strengths $u$ shown in the legend. The binding energy passes through a maximum as the width is increased, after which it falls back to zero.}
		\label{fig:sqSize}
	\end{center}
\end{figure}

Fig.~\ref{fig:sqSize} shows the binding energy of the subgap states as a function of potential width when the potential strength is held constant (fixed $u$, variable $L$). The figure indicates resonant behavior: for a given field strength, the binding energy passes through a maximum as $L$ is increased from zero.

The existence of a maximum binding energy follows from the limiting cases $L \ll R$ and $L \gg R$. The energy of the bound state is the solution of the secular equation, Eq.~(\ref{eq:secular}). When $L$ is small, the binding energy increases in proportion to $u^4$. When $L$ is large, $\Lambda(\varepsilon)$ is proportional to $L$, and Eq.~(\ref{eq:secular}) effectively becomes $\det \mathbb{G}_0 \Gamma_0 = 0$, which gives $\varepsilon = \Delta_0$. Corrections are of order $R/L$. Since the binding energy increases initially and approaches zero in the limit $L \gg R$, it must pass through a maximum. This leads to the peaks shown in Fig.~\ref{fig:sqSize}.

The decrease in binding energy with increasing length should not be interpreted as the uniform field limit of the square well model. The results of Fig.~\ref{fig:sqSize} were obtained by assuming the square well was smaller than the wavelength of a particle in the lowest band, which is not satisfied in a uniform field.

These two figures indicate that a strong electric field applied over a small region of the nanotube is required to create localized states states far from the band edge.

\section{Conclusion}\label{sec:final}

In this paper, we have considered the effects of nonuniform \tef{}s on the electronic structure of semiconducting carbon nanotubes by using the linear Dirac Hamiltonian for states near the Fermi surface. We considered fields that vary slowly over distances comparable to the interatomic spacing, but rapidly over distances comparable to the wavelength of the envelope functions. This allows us to ignore intervalley scattering even though the effective potentials in the long-wavelength Hamiltonian change discontinuously.

We studied two effective potentials: a delta function and a square well. In both models, the potential leads to the formation of localized states inside the free nanotube band gap for arbitrarily weak fields. For weak fields, the energy of the subgap states moves away from the band edge in proportion to the fourth power of the electric field strength. When the range of the potential is fixed, the binding energy of the subgap states exhibits crossover behavior. The states lie close to the unperturbed band edge until the field strength exceeds a threshold value, after which they rapidly approach the center of the band. In the square well model, the binding energy of the subgap state passes through a maximum as the width of the well is increased. In both models, a strong electric field applied to a small region of the nanotube is required to create subgap states far from the band edge.

The subgap states are exponentially localized around the center of the potential. As such a single subgap state is not likely to affect transport along the nanotube axis. However, these states might be important in processes involving transport perpendicular to the nanotube axis, such as scanning tunneling microscopy experiments. The effect on STM experiments designed to measure the nanotube band gap should be negligible, however. If the tip of the scanning probe just touches the nanotube, which just touches the substrate into which electrons will flow, then the electric field in the nanotube is roughly equal to the potential drop across its diameter divided by the static dielectric constant.  In order for electrons to tunnel into the nanotube, the potential due to the STM tip must be on the order of the band gap: $e \Delta V \approx 2 \hbar v_f / 3R$. This gives $u \approx 1/3$. The subgap states at this field strength are virtually indistinguishable from the free nanotube band edge in either model.

One might also consider several of these subgap states evenly spaced along the nanotube axis. This array could be used to control tunneling through the nanotube for charge carriers with energies inside the band gap. By tuning the energy of the subgap states, one could control tunneling across the nanotube from a reservoir of charge carriers at some energy $\varepsilon < \Delta_0$. Subgap states created by an external potential could also be important in exciton dynamics. They might provide specific sites along the tube for exciton localization and recombination.

Experiments to explore this effect would require the generation of large electric fields --- tens of volts per nanometer --- in a relatively small region along the nanotube axis. One experimental signature of the subgap states would be an increase in tunneling current perpendicular to the nanotube axis with no change in transport along the tube axis at the same potential bias.

The Green function approach we used is not limited to carbon nanotubes. The theory of graphene nanoribbons is similar to that of carbon nanotubes. The major difference is that periodic boundary conditions are replaced by hard wall boundary conditions. A uniform static electric field perpendicular to the edge of a nanoribbon applied over a small region would lead to a potential similar to Eq.~(\ref{eq:potential}). However, for nanoribbons, the applied field would mix all bands since the potential is linear instead of periodic. The methods described in this paper might be useful in spite of the differences.

It would also be desirable to compare our predictions with the results of numerical simulations. However, it seems that the conditions in which our effective potentials accurately approximate the physical field make numerical simulation difficult. The potential breaks translation invariance and extends over a large number of unit cells, but a very small fraction of the length of the nanotube. As a result, the number of atoms and unit cells required would make a tight-binding simulation or density functional theory challenging.

This work was supported by DOE grant No. DE-FG02-ER0145118, and by the NSF under grant DMR-00-79909.

\appendix

\section{Comparison of Analytic and Numerical Results for Uniform STEF}\label{app:table}

Table~\ref{tab:compare} compares analytic and numerical predictions for the band gap response of a semiconducting carbon nanotube in a uniform \tef. The first two columns give the chiral indices and radius of the nanotube analyzed. The third column gives the critical field strength predicted by Novikov and Levitov:\cite{nov02}
\begin{align*}
	E_c &\approx 0.62 \cdot  \hbar v_F / e R^2  \\
		&\approx \dfrac{ 205 }{ m^2 + mn + n^2 } \ \mbox{V/nm}.
\end{align*}
The fourth column gives the approximate value of the threshold electric field strength from numerical simulations (the field strength at which the band gap started to decrease linearly with increasing field strength, interpolated from the graphs given in the references cited), and the fifth column gives the reference from which this value was taken. All numerical simulations were for semiconducting zig-zag nanotubes. The electric field strengths do not include depolarization effects. The values represent the field inside the nanotube, not the applied electric field.

\begin{table}[bh]
\begin{ruledtabular}
\begin{tabular}{ccccc}
	Tube & Radius & Critical Field & Threshold & Reference \\
	$(m,n)$ & (nm) & (V/nm) & (V/nm) & \\
	\hline
	(10,0) & 0.4 & 2 & 3.3 & Ref.~\onlinecite{cho02} \\
	(8,0) & 0.3 & 3.2 & 4 & Ref.~\onlinecite{pach05} \\
	(11,0) & 0.4 & 1.7 & 2 & Ref.~\onlinecite{pach05} \\
	(16,0) & 0.6 & 0.8 & 1.2 & Ref.~\onlinecite{li06} \\
\end{tabular}
\end{ruledtabular}
\caption{Comparison of the critical field strength predicted by Novikov and Levitov and the threshold field reported in numerical tight binding calculations. The critical field from the linear Dirac theory provides a physical explanation for the crossover behavior observed in simulations.}
\label{tab:compare}
\end{table}

The Dirac Hamiltonian is the linear $\mathbf{k \cdot p}$ approximation of a tight-binding Hamiltonian for carbon nanotubes, so the results of the two theories should agree, at least qualitatively. Novikov and Levitov acknowledge that higher-order terms break the symmetry that protects the band gap,\cite{nov02} so there is no true critical behavior. However, the critical field strength of the Dirac theory provides an intuitive explanation for the crossover behavior reported in numerical simulations. The low-energy Dirac theory captures the essential physics of carbon nanotubes necessary for a qualitative description of the response to an external electric field.

\section{No Dependence on Chiral Angle}\label{app:noChiral}

Here, we show that the density of states and the energy eigenvalues of the subgap bound states do not depend on the chiral angle.

The Hamiltonian for a nanotube in a nonuniform \tef{} can be written
\begin{equation}
	\mathcal{H}(\theta) = e^{-i \sigma_z \theta / 2} [ ( \sigma_y \Delta_m - i R \sigma_x \pd{z}  ) \, \delta_{m,n} + ( u / 2 ) \, f(z) \, \delta_{m,n\pm1} ] \, e^{i \sigma_z \theta / 2}. \label{eq:hChiral}
\end{equation}
This is equivalent to Eq.~(\ref{eq:hRed}). (Since the potential is a scalar, $e^{-i \sigma_z \theta / 2} \, V_{m,n}(z) \, e^{i \sigma_z \theta / 2} = V_{m,n}(z)$.) We will refer to the Green functions derived from this Hamiltonian as ``chiral.'' Those studied in the main body of the text do not depend on the chiral angle, so we will refer to them as ``achiral.''

The equation satisfied by the chiral Green functions is
\begin{equation}
	[ \varepsilon \, \delta_{m,n} - \mathcal{H}_{m,n}(\theta) ] \, \mathfrak{S}_m(\theta) = \delta(z). \label{eq:gChiral}
\end{equation}

Since $\varepsilon$ is also a scalar, we may write
\begin{equation}
	e^{-i \sigma_z \theta / 2} \, \left( \varepsilon \, \delta_{m,n} - \mathcal{H}_{m,n} \right) \, e^{i \sigma_z \theta / 2} \, \mathfrak{S}_m(\theta) = \delta(z).
\end{equation}

Multiplying on the left by $e^{i \sigma_z \theta / 2}$ and on the right by $e^{-i \sigma_z \theta / 2}$, we obtain an achiral operator acting on a transformed Green function. The transformed Green function is the inverse of an operator that does not depend on the chiral angle, and so it too must be independent of chiral angle. The equation satisfied by the achiral Green functions is:
\begin{equation}
	\left( \varepsilon \delta_{m,n} - \mathcal{H}_{m,n} \right) \, \mathfrak{S}_m = \delta(z).
\end{equation}
The chiral and achiral Green functions are related by a unitary transformation:
\begin{equation}
	\mathfrak{S}_m = e^{i \sigma_z \theta / 2} \, \mathfrak{S}_m(\theta) \, e^{-i \sigma_z \theta / 2}. \label{eq:chiral}
\end{equation}

The results of the main body of the paper apply to the achiral Green functions. Next, we show that the bound state energies and density of states derived from achiral Green function $\mathfrak{S}_m$ are the same as those derived from the chiral Green functions. This implies that the density of states and bound state energies do not depend on the chiral angle. 

The density of states depends only on the trace of the Green function. Since the trace of a matrix is invariant under unitary transformations, the density of states derived from $\mathfrak{S}_m$ and $\mathfrak{S}_m(\theta)$ are identical.

The energies of the bound states of the $m$-th band are the poles of the chiral Green function, which occur when the determinant of the denominator in Eq.~(\ref{eq:genSol0}) vanishes:
\begin{equation}
	\det \left\{ \openone - \mathbb{G}_m \Gamma_m \right\} = 0. \label{eq:detZero}
\end{equation}

The matrix $\Gamma_m$ is a linear combination of two achiral Green functions:
\begin{equation}
	\Gamma_m = \alpha \mathbb{G}_{m+1} + \beta \mathbb{G}_{m-1}.
\end{equation}
As a result, the relation between $\Gamma_m$ and the chiral matrix $\Gamma_m(\theta)$ is identical to that between the Green functions, given in Eq.~(\ref{eq:chiral}).

Since the identity matrix is invariant under unitary transformations, Eq.~(\ref{eq:detZero}) can be expressed in terms of the chiral Green functions as
\begin{equation}
	\det \left\{ e^{i \sigma_z \theta / 2} \, \left[ \openone - \mathbb{G}_m(\theta) \Gamma_m(\theta) \right] \, e^{-i \sigma_z \theta / 2} \right\} = 0.
\end{equation}
The determinant of a matrix is invariant under unitary transformations. Therefore, the determinants of the chiral the achiral Hamiltonian are equal, and lead to the same electronic spectra.

To summarize, neither the density of states nor the energies of the subgap states depends on the chiral angle of the nanotube for the models studied in this paper. This is consistent with known results for a free nanotube or a nanotube in a uniform \tef. In these models, there is no dependence on the chiral angle when the effective Hamiltonian includes only terms linear in $\pd{z}$ and $\pd{\phi}$ (or $q$ and $\Delta_m$ in Fourier space).

There are two ways in which a dependence on the chiral angle might be introduced. First, one could introduce the quadratic terms of the $\mathbf{k \cdot p}$ approximation to the effective Hamiltonian with a scalar potential. Trigonal warping cannot be removed with a unitary transformation and leads to a dispersion relation that depends on the chiral angle. A second possibility is to introduce a potential that mixes states on different sublattices. Even if the effective Hamiltonian is linear in $\pd{z}$ and $\pd{\phi}$, the potential will introduce a term proportional to $\sigma_x$, $\sigma_y$, or some linear combination of the two. In this case, the chiral Hamiltonian $\mathcal{H}(\theta)$ is no longer related to an achiral operator by a unitary transformation, and the eigenvalue spectrum will depend on the chiral angle.

\section{Derivation of Green Functions}\label{app:gFunc}

Here, we derive the achiral Green functions for the unperturbed Hamiltonian. As discussed in App.~\ref{app:noChiral}, the bound state energies and the density of states do not depend on the chiral angle, so we ignore it to simplify the calculation.

Because the free nanotube Hamiltonian is static and invariant under translations along the nanotube axis, the Green functions can only depend on the difference between coordinates: $\mathbb{G}(z,z') = \mathbb{G}_m(z-z')$. We set the $z'$ to zero when solving for the Green function. The equation satisfied by $\mathbb{G}_m(z)$ is
\begin{equation}
	\left( \varepsilon - \sigma_y \, \Delta_m + i R \sigma_x \pd{z} \right) \mathbb{G}_m(\varepsilon,z) = \delta(z). \label{eq:greenZero}
\end{equation}

This linear differential equation can be solved by Fourier transformation:
\begin{equation}
	\mathbb{G}_m(\varepsilon,z) = \int_{-\infty}^{\infty} \dfrac{ \d q }{ 2 \pi } \,e^{ - i q z } \, \mathbb{G}_m(\varepsilon,q).
\end{equation}

Using this expression for $\mathbb{G}_m(\varepsilon,z)$ in Eq.~(\ref{eq:greenZero}), the solution is
\begin{equation}
	\mathbb{G}_m(\varepsilon,q) = \dfrac{ \varepsilon + \sigma_y \Delta_m + \sigma_x qR }{ \varepsilon^2 - (qR)^2 - (\Delta_m)^2 + i \eta }.
\end{equation}
A convergence factor $i \eta$ has been added to the denominator to push the poles of the Green function into the appropriate quadrants of the complex plane. This Green function will satisfy boundary conditions for outgoing waves as $z$ approaches $\pm \infty$.

Since the potentials considered in this paper break translation symmetry along the nanotube axis, the real space Green function is easier to use in calculations. Inverting the Fourier transform for $q$ gives
\begin{align}
	\mathbb{G}_m(\varepsilon,z) &= \int_{-\infty}^{\infty} \, \dfrac{ \d q }{ 2 \pi } \, e^{ i q z } \, \mathbb{G}_m(\varepsilon,q) \notag \\
		&= \dfrac{ \varepsilon + \sigma_y \Delta_m + \mbox{sgn}(z-z') \, \sigma_x \xi_m(\varepsilon) }{ 2 i R \xi_m(\varepsilon)  } \, e^{ i \xi_m(\varepsilon) \, |z-z'| / R  }.
\end{align}
where $\xi_m(\varepsilon) = \sqrt{ \varepsilon^2 - \Delta_{m}^{2} }$. This is equivalent to the Green function given in Eq.~(\ref{eq:green}).

\section{Justification for Ignoring Transitions to Bands with $|\Delta m| > 1$}\label{app:neighbors}

Here, we argue that the error introduced by ignoring all terms in the Dyson series that involve transitions to bands other than the two nearest neighbors is small for energies near the band edge.

To keep the expressions simple, we consider the full Green function for $z = z' = 0$ and suppress dependence on $z$, $z'$, and $\varepsilon$. In the delta function model, the series for the full Green function in the lowest band is
\begin{align}
	\mathfrak{S}_0 = \mathbb{G}_0 &+ u^2 \ \mathbb{G}_0 \left( \mathbb{G}_{+1} + \mathbb{G}_{-1} \right) \, \mathbb{G}_0 \notag \\
		&+ u^4 \ \mathbb{G}_0 \left( \mathbb{G}_{+1} + \mathbb{G}_{-1} \right) \, \mathbb{G}_0 \left( \mathbb{G}_{+1} + \mathbb{G}_{-1} \right) \, \mathbb{G}_0 \\
		&+ u^4 \ \mathbb{G}_0 \, \left( \mathbb{G}_{+1} \mathbb{G}_{+2} \mathbb{G}_{+1} + \mathbb{G}_{-1} \mathbb{G}_{-2} \mathbb{G}_{-1} \right) \, \mathbb{G}_0 + \mathcal{O}(u^6). \notag
\end{align}

All terms of the infinite series involving only transitions to the two neighboring bands can be generated from a single term, which we call $\Sigma$:
\begin{equation}
	\Sigma = u^2 \ \left( \mathbb{G}_{+1} + \mathbb{G}_{-1} \right) \, \mathbb{G}_0.
\end{equation}

The $\mathcal{O}(u^{2n})$ term of the series can be expressed as $(\Sigma)^n$ plus other terms that involve Green functions from bands other than $m = 0, \pm1$. For instance, the $\mathcal{O}(u^4)$ terms can be expressed
\begin{equation}
	\mathbb{G}_0 \, \Sigma^2 + u^4 \ \mathbb{G}_0 \, \left( \mathbb{G}_{+1} \mathbb{G}_{+2} \mathbb{G}_{+1} + \mathbb{G}_{-1} \mathbb{G}_{-2} \mathbb{G}_{-1} \right) \, \mathbb{G}_0.
\end{equation}

In the delta function model, the terms of the series are products of the free nanotube Green functions in Eq.~(\ref{eq:green}) evaluated at $z = z' = 0$:
\begin{equation}
	\mathbb{G}_n = - \dfrac{ i }{ 2 R^2 } \, \dfrac{ \varepsilon + \sigma_y \Delta_n }{ \sqrt{ \varepsilon^2 - (\Delta_n)^2 } }.
\end{equation}
Near the band edge, $\varepsilon \approx \Delta_0$. Therefore, the denominator of $\mathbb{G}_0$ is very small, but the denominator of $\mathbb{G}_n$ with $n \neq 0$ is of order unity. As a result, the largest terms of the series at any order are those with the most factors of $\mathbb{G}_0$ in the product --- i.e. the terms generated by $(\Sigma)^n$.

As an example, we compare the two $\mathcal{O}(u^4)$ terms
\begin{equation}
	\mathbb{G}_0 \mathbb{G}_{+1}\mathbb{G}_0 \mathbb{G}_{+1} \mathbb{G}_0 \sim \left( \dfrac{ 1 }{ \sqrt{ \varepsilon^2 - (\Delta_0)^2 } } \right)^3 \left( \dfrac{ 1 }{ \sqrt{ \varepsilon^2 - (\Delta_{+1})^2 } } \right)^2 \label{eq:bigTerms}
\end{equation}
and
\begin{equation}
	\mathbb{G}_0 \mathbb{G}_{+1}\mathbb{G}_{+2} \mathbb{G}_{+1} \mathbb{G}_0 \sim \left( \dfrac{ 1 }{ \sqrt{ \varepsilon^2 - (\Delta_0)^2 } } \right)^2 \left( \dfrac{ 1 }{ \sqrt{ \varepsilon^2 - (\Delta_{+1})^2 } } \right)^2 \left( \dfrac{ 1 }{ \sqrt{ \varepsilon^2 - (\Delta_{+2})^2 } } \right).\label{eq:smallTerms}
\end{equation}
We have ignored all common factors, as well as the numerators, which are matrix products of terms of order unity. Eq.~(\ref{eq:bigTerms}) is generated by $(\Sigma)^2$. It is greater than the second term by a factor of $ \sqrt{ \varepsilon^2 - (\Delta_{+2})^2 } / \sqrt{ \varepsilon^2 - (\Delta_{0})^2 } $, which is large when $\varepsilon$ is near the band edge. For instance, if $|\varepsilon|/|\Delta_0| = 0.1$, the first term is large than the second by a factor of roughly 16.

Thus, the error introduced by discarding all terms except those generated by $\Sigma$ is small when the energy is near the band edge. We approximate the full Green function by a partial summation
\begin{equation}
	\mathfrak{S}_0 \approx \mathbb{G}_0 \sum_{n=0}^{\infty} (\Sigma)^n.
\end{equation}

Multiplying by the right by $\Sigma$, we have
\begin{equation}
	\mathfrak{S}_0 \, \Sigma = \mathbb{G}_0 \sum_{n=0}^{\infty} (\Sigma)^{n+1} = \mathfrak{S}_0 - \mathbb{G}_0.
\end{equation}

This can be solved to give
\begin{equation}
	\mathfrak{S}_0 = \mathbb{G}_0 \, ( \openone - \Sigma )^{-1}.
\end{equation}
(Formally, this series only converges when the eigenvalues of $\Sigma$ are less than one.\cite{matt92,don92})

Eq.~(\ref{eq:genSol}) gives the most general solution, with $z \neq z'$, and its derivation closely parallels the steps in this appendix.  If $z = z'$, Eq.~(\ref{eq:genSol}) reduces to $\mathbb{G}_0 \, ( \openone - \Sigma )^{-1}$.

\section{Bound State for Any Finite Potential}\label{app:noCrit}

Here, we show that subgap states exist for arbitrarily weak fields in the delta function model of Sec.~\ref{sec:delta}.

If there are poles inside the band gap, then the denominator of Eq.~(\ref{eq:genSol}) must vanish for some energy in the range $0 < \varepsilon < \Delta_0$. To simplify the expressions that follow, we define $\zeta = \varepsilon / \Delta_0$ as the ratio of the particle energy to the unperturbed band gap, and $\chi = u L / 4 R$ as a dimensionless measure of the potential strength. Poles will be solutions of the equation
\begin{equation}
	0 =\det \left[ \openone - \chi^2 \, \dfrac{ \zeta + \sigma_y }{\sqrt{1 - \zeta^2}} \left( \dfrac{ \zeta + \sigma_y \left( 1 + 1 / \Delta_0 \right) }{ \sqrt{ \left( 1 + 1 / \Delta_0 \right)^2 - \zeta^2 } } + \dfrac{ \zeta + \sigma_y \left( 1 - 1 / \Delta_0 \right) }{ \sqrt{ \left( 1 - 1 / \Delta_0 \right)^2 - \zeta^2 } } \right) \right].
\end{equation}

After multiplying out all the terms, the resulting product is a matrix of the form $\mathbb{M} = A(\zeta) \, \openone + B(\zeta) \, \sigma_y$. The determinant of such a matrix is $A(\zeta)^2 - B(\zeta)^2$. In this case, substituting $\Delta_0 = 1/3$ gives
\begin{equation}
	A(\zeta) = 1 - \chi^2 \dfrac{\left( \zeta^2 + 4 \right) \sqrt{ 4 - \zeta^2 } + \left( \zeta^2 - 2 \right) \sqrt{ 16 - \zeta^2 }}{\sqrt{ \left( 1 - \zeta^2 \right) \left( 16 - \zeta^2 \right) \left( 4 - \zeta^2 \right) }},
\end{equation}

\begin{equation}
	B(\zeta) = \chi^2 \dfrac{ \sqrt{ 16 - \zeta^2 } - 5 \sqrt{ 4 - \zeta^2 } }{\sqrt{ \left( 1 - \zeta^2 \right) \left( 16 - \zeta^2 \right) \left( 4 - \zeta^2 \right) }} \, \zeta.
\end{equation}

If there are bound states inside the band gap, then we expect their energy to be close to the band edge when the potential is small --- i.e. if $u L \ll R $, then $\zeta \approx 1$. Letting $\zeta = 1 - \eta$ then Taylor expanding $A(\zeta)^2 - B(\zeta)^2 = 0$ to $\mathcal{O}(\eta)$ gives
\begin{equation}
	0 = \left( 90 - 36 \sqrt{5} \, \left(3 - \sqrt{5} \right) \, \chi^4 \right) \, \eta  - 30 \sqrt{ 6 } \, \left( \sqrt{ 5 } - 1 \right) \chi^2 \, \sqrt{\eta} + \mathcal{O}(\eta^{3/2}).
\end{equation}

Neglecting the terms of order $\eta^{3/2}$ and higher, this equation has two solutions. One of these is $\eta = 0$. This solution corresponds to $\varepsilon = \Delta_0$, which is not a pole of the full Green function in Eq.~(\ref{eq:genSol}). The other solution is
\begin{equation}
	\eta = \left( \dfrac{ \chi^2 }{ r - s \, \chi^4 } \right)^2
\end{equation}
where $r \approx 0.991$ and $s \approx 0.677$. Thus, any nonzero applied field produces a bound state inside the band gap, and the binding energy of this state scales with the fourth power of the potential strength when the potential is small. The singularity in this expression is a consequence of truncating the Taylor series. The expression diverges when $\chi \approx 1.01$, but the expression was derived assuming $\chi \ll 1$. Fig.~\ref{fig:delta} shows that the energy of the subgap states has no singularities.


\begin{thebibliography}{19}
\expandafter\ifx\csname natexlab\endcsname\relax\def\natexlab#1{#1}\fi
\expandafter\ifx\csname bibnamefont\endcsname\relax
  \def\bibnamefont#1{#1}\fi
\expandafter\ifx\csname bibfnamefont\endcsname\relax
  \def\bibfnamefont#1{#1}\fi
\expandafter\ifx\csname citenamefont\endcsname\relax
  \def\citenamefont#1{#1}\fi
\expandafter\ifx\csname url\endcsname\relax
  \def\url#1{\texttt{#1}}\fi
\expandafter\ifx\csname urlprefix\endcsname\relax\def\urlprefix{URL }\fi
\providecommand{\bibinfo}[2]{#2}
\providecommand{\eprint}[2][]{\url{#2}}

\bibitem[{\citenamefont{{Kane} and {Mele}}(1997)}]{kane97}
\bibinfo{author}{\bibfnamefont{C.~L.} \bibnamefont{{Kane}}} \bibnamefont{and}
  \bibinfo{author}{\bibfnamefont{E.~J.} \bibnamefont{{Mele}}},
  \bibinfo{journal}{Physical Review Letters} \textbf{\bibinfo{volume}{78}},
  \bibinfo{pages}{1932} (\bibinfo{year}{1997}).

\bibitem[{\citenamefont{Ando}(2004)}]{ando02}
\bibinfo{author}{\bibfnamefont{T.}~\bibnamefont{Ando}},
  \bibinfo{journal}{Journal of the Physical Society of Japan}
  \textbf{\bibinfo{volume}{73}}, \bibinfo{pages}{3351} (\bibinfo{year}{2004}).

\bibitem[{\citenamefont{Chen et~al.}(2004)\citenamefont{Chen, Lee, and
  Clark}}]{chen04}
\bibinfo{author}{\bibfnamefont{C.-W.} \bibnamefont{Chen}},
  \bibinfo{author}{\bibfnamefont{M.-H.} \bibnamefont{Lee}}, \bibnamefont{and}
  \bibinfo{author}{\bibfnamefont{S.~J.} \bibnamefont{Clark}},
  \bibinfo{journal}{Nanotechnology} \textbf{\bibinfo{volume}{15}},
  \bibinfo{pages}{1837} (\bibinfo{year}{2004}).

\bibitem[{\citenamefont{Khoo et~al.}(2004)\citenamefont{Khoo, Mazzoni, and
  Louie}}]{khoo04}
\bibinfo{author}{\bibfnamefont{K.~H.} \bibnamefont{Khoo}},
  \bibinfo{author}{\bibfnamefont{M.~S.~C.} \bibnamefont{Mazzoni}},
  \bibnamefont{and} \bibinfo{author}{\bibfnamefont{S.~G.} \bibnamefont{Louie}},
  \bibinfo{journal}{Physical Review B} \textbf{\bibinfo{volume}{69}},
  \bibinfo{pages}{201401(R)} (\bibinfo{year}{2004}).

\bibitem[{\citenamefont{Ishigami et~al.}(2005)\citenamefont{Ishigami, Sau,
  Aloni, Cohen, and Zettl}}]{ishi05}
\bibinfo{author}{\bibfnamefont{M.}~\bibnamefont{Ishigami}},
  \bibinfo{author}{\bibfnamefont{J.~D.} \bibnamefont{Sau}},
  \bibinfo{author}{\bibfnamefont{S.}~\bibnamefont{Aloni}},
  \bibinfo{author}{\bibfnamefont{M.~L.} \bibnamefont{Cohen}}, \bibnamefont{and}
  \bibinfo{author}{\bibfnamefont{A.}~\bibnamefont{Zettl}},
  \bibinfo{journal}{Physical Review Letters} \textbf{\bibinfo{volume}{94}},
  \bibinfo{eid}{056804} (\bibinfo{year}{2005}),
  \urlprefix\url{http://link.aps.org/abstract/PRL/v94/e056804}.

\bibitem[{\citenamefont{{O'Keeffe} et~al.}(2002)\citenamefont{{O'Keeffe},
  {Wei}, and {Cho}}}]{cho02}
\bibinfo{author}{\bibfnamefont{J.}~\bibnamefont{{O'Keeffe}}},
  \bibinfo{author}{\bibfnamefont{C.}~\bibnamefont{{Wei}}}, \bibnamefont{and}
  \bibinfo{author}{\bibfnamefont{K.}~\bibnamefont{{Cho}}},
  \bibinfo{journal}{Applied Physics Letters} \textbf{\bibinfo{volume}{80}},
  \bibinfo{pages}{676} (\bibinfo{year}{2002}).

\bibitem[{\citenamefont{Novikov and Levitov}(2002)}]{nov02}
\bibinfo{author}{\bibfnamefont{D.~S.} \bibnamefont{Novikov}} \bibnamefont{and}
  \bibinfo{author}{\bibfnamefont{L.~S.} \bibnamefont{Levitov}},
  \emph{\bibinfo{title}{Electron properties of carbon nanotubes in the field
  effect regime}} (\bibinfo{year}{2002}),
  \urlprefix\url{http://arxiv.org/abs/cond-mat/0204499}.
  
\bibitem[{\citenamefont{{Novikov} and {Levitov}}(2006)}]{nov06}
\bibinfo{author}{\bibfnamefont{D.~S.} \bibnamefont{{Novikov}}}
  \bibnamefont{and} \bibinfo{author}{\bibfnamefont{L.~S.}
  \bibnamefont{{Levitov}}}, \bibinfo{journal}{Physical Review Letters}
  \textbf{\bibinfo{volume}{96}}, \bibinfo{pages}{036402}
  (\bibinfo{year}{2006}).

\bibitem[{\citenamefont{Li et~al.}(2003)\citenamefont{Li, Rotkin, and
  Ravaioli}}]{rot03}
\bibinfo{author}{\bibfnamefont{Y.}~\bibnamefont{Li}},
  \bibinfo{author}{\bibfnamefont{S.}~\bibnamefont{Rotkin}}, \bibnamefont{and}
  \bibinfo{author}{\bibfnamefont{U.}~\bibnamefont{Ravaioli}},
  \bibinfo{journal}{Nano Letters} \textbf{\bibinfo{volume}{3}},
  \bibinfo{pages}{183} (\bibinfo{year}{2003}).

\bibitem[{\citenamefont{Pacheco et~al.}(2005)\citenamefont{Pacheco, Barticevic,
  Rocha, and Latg{\'e}}}]{pach05}
\bibinfo{author}{\bibfnamefont{M.}~\bibnamefont{Pacheco}},
  \bibinfo{author}{\bibfnamefont{Z.}~\bibnamefont{Barticevic}},
  \bibinfo{author}{\bibfnamefont{C.~G.} \bibnamefont{Rocha}}, \bibnamefont{and}
  \bibinfo{author}{\bibfnamefont{A.}~\bibnamefont{Latg{\'e}}},
  \bibinfo{journal}{Journal of Physics: Condesned Matter}
  \textbf{\bibinfo{volume}{17}}, \bibinfo{pages}{5839} (\bibinfo{year}{2005}).

\bibitem[{\citenamefont{Li and Lin}(2006)}]{li06}
\bibinfo{author}{\bibfnamefont{T.~S.} \bibnamefont{Li}} \bibnamefont{and}
  \bibinfo{author}{\bibfnamefont{M.~F.} \bibnamefont{Lin}},
  \bibinfo{journal}{Physical Review B} \textbf{\bibinfo{volume}{73}},
  \bibinfo{pages}{075432} (\bibinfo{year}{2006}).

\bibitem[{\citenamefont{Gunlycke et~al.}(2006)\citenamefont{Gunlycke, Lambert,
  Bailey, Pettifor, Briggs, and Jefferson}}]{gun06}
\bibinfo{author}{\bibfnamefont{D.}~\bibnamefont{Gunlycke}},
  \bibinfo{author}{\bibfnamefont{C.~J.} \bibnamefont{Lambert}},
  \bibinfo{author}{\bibfnamefont{S.~W.~D.} \bibnamefont{Bailey}},
  \bibinfo{author}{\bibfnamefont{D.~G.} \bibnamefont{Pettifor}},
  \bibinfo{author}{\bibfnamefont{G.~A.~D.} \bibnamefont{Briggs}},
  \bibnamefont{and} \bibinfo{author}{\bibfnamefont{J.~H.}
  \bibnamefont{Jefferson}}, \bibinfo{journal}{Europhysics Letters}
  \textbf{\bibinfo{volume}{73}}, \bibinfo{pages}{759} (\bibinfo{year}{2006}).

\bibitem[{\citenamefont{{Wallace}}(1947)}]{wal47}
\bibinfo{author}{\bibfnamefont{P.~R.} \bibnamefont{{Wallace}}},
  \bibinfo{journal}{Physical Review} \textbf{\bibinfo{volume}{71}},
  \bibinfo{pages}{622} (\bibinfo{year}{1947}).

\bibitem[{\citenamefont{Semenoff}(1984)}]{sem84}
\bibinfo{author}{\bibfnamefont{G.~W.} \bibnamefont{Semenoff}},
  \bibinfo{journal}{Physical Review Letters} \textbf{\bibinfo{volume}{53}},
  \bibinfo{pages}{2449} (\bibinfo{year}{1984}).

\bibitem[{\citenamefont{DiVincenzo and Mele }(1984)}]{mele84}
\bibinfo{author}{\bibfnamefont{D.~P.} \bibnamefont{DiVincenzo}} \bibnamefont{and}
  \bibinfo{author}{\bibfnamefont{E.~J.} \bibnamefont{Mele}},
  \bibinfo{journal}{Physical Review B} \textbf{\bibinfo{volume}{29}},
  \bibinfo{pages}{1685} (\bibinfo{year}{1984}).

\bibitem[{\citenamefont{Benedict et~al.}(1995)\citenamefont{Benedict, Louie,
  and Cohen}}]{ben95}
\bibinfo{author}{\bibfnamefont{L.~X.} \bibnamefont{Benedict}},
  \bibinfo{author}{\bibfnamefont{S.~G.} \bibnamefont{Louie}}, \bibnamefont{and}
  \bibinfo{author}{\bibfnamefont{M.~L.} \bibnamefont{Cohen}},
  \bibinfo{journal}{Physical Review B} \textbf{\bibinfo{volume}{52}},
  \bibinfo{pages}{8541} (\bibinfo{year}{1995}).

\bibitem[{\citenamefont{Mattuck}(1992)}]{matt92}
\bibinfo{author}{\bibfnamefont{R.~D.} \bibnamefont{Mattuck}},
  \emph{\bibinfo{title}{A Guide to Feynman Diagrams in the Many-Body Problem}}
  (\bibinfo{publisher}{Dover}, \bibinfo{year}{1992}), \bibinfo{edition}{2nd}
  ed.

\bibitem[{\citenamefont{Doniach and Sondheimer}(1974)}]{don92}
\bibinfo{author}{\bibfnamefont{S.}~\bibnamefont{Doniach}} \bibnamefont{and}
  \bibinfo{author}{\bibfnamefont{E.}~\bibnamefont{Sondheimer}},
  \emph{\bibinfo{title}{Green's Functions for Solid State Physicists}}
  (\bibinfo{publisher}{W.A. Benjamin, Inc.}, \bibinfo{year}{1974}),
  \bibinfo{edition}{1st} ed.

\end{thebibliography}
\end{document}